\documentclass[final,5p,times]{elsarticle} 
\usepackage{amssymb} 
\usepackage{amsthm}
\usepackage{amsmath}
\usepackage{lineno,hyperref,graphicx}\hypersetup{colorlinks = true}
\modulolinenumbers[5]
\DeclareGraphicsExtensions{.eps,.png,.jpg,.pdf}

\begin{document}
\begin{frontmatter}
\title{WAVE: Machine Learning for Full-Waveform Time-Of-Flight Detectors}

\author[UltralyticsLLC]{Glenn R. Jocher}\corref{cor1}\ead{glenn.jocher@ultralytics.com}
\author[UH]{Kurtis Nishimura}\ead{kurtis.nishimura@ultralytics.com}
\author[UH]{John Koblanski}\ead{johnk2@hawaii.edu}
\author[LLNL]{Viacheslav A. Li}\ead{vli2@hawaii.edu}

\address[UltralyticsLLC]{Ultralytics LLC, Arlington, VA, USA}
\address[UH]{Department of Physics and Astronomy, University of Hawaii, Honolulu, HI 96822, USA}
\address[LLNL]{Lawrence Livermore National Laboratory, CA 94550}
\cortext[cor1]{Corresponding author}

\begin{abstract}
We propose a \textbf{WA}veform \textbf{V}ector \textbf{E}xploitation (\textbf{WAVE}) deep neural network for full-waveform Time-Of-Flight (TOF) physics detectors, and evaluate its performance against traditional reconstruction techniques via Monte Carlo study of a small plastic-scintillator scatter camera. Ultralytics LLC (\url{https://www.ultralytics.com}) provides WAVE freely under the open source GPL-3.0 license at \url{https://github.com/ultralytics/wave}.
\end{abstract}


\end{frontmatter}
\tableofcontents

\section{Introduction}

\begin{figure}[!htbp]\centering
\includegraphics[width=.95\linewidth]{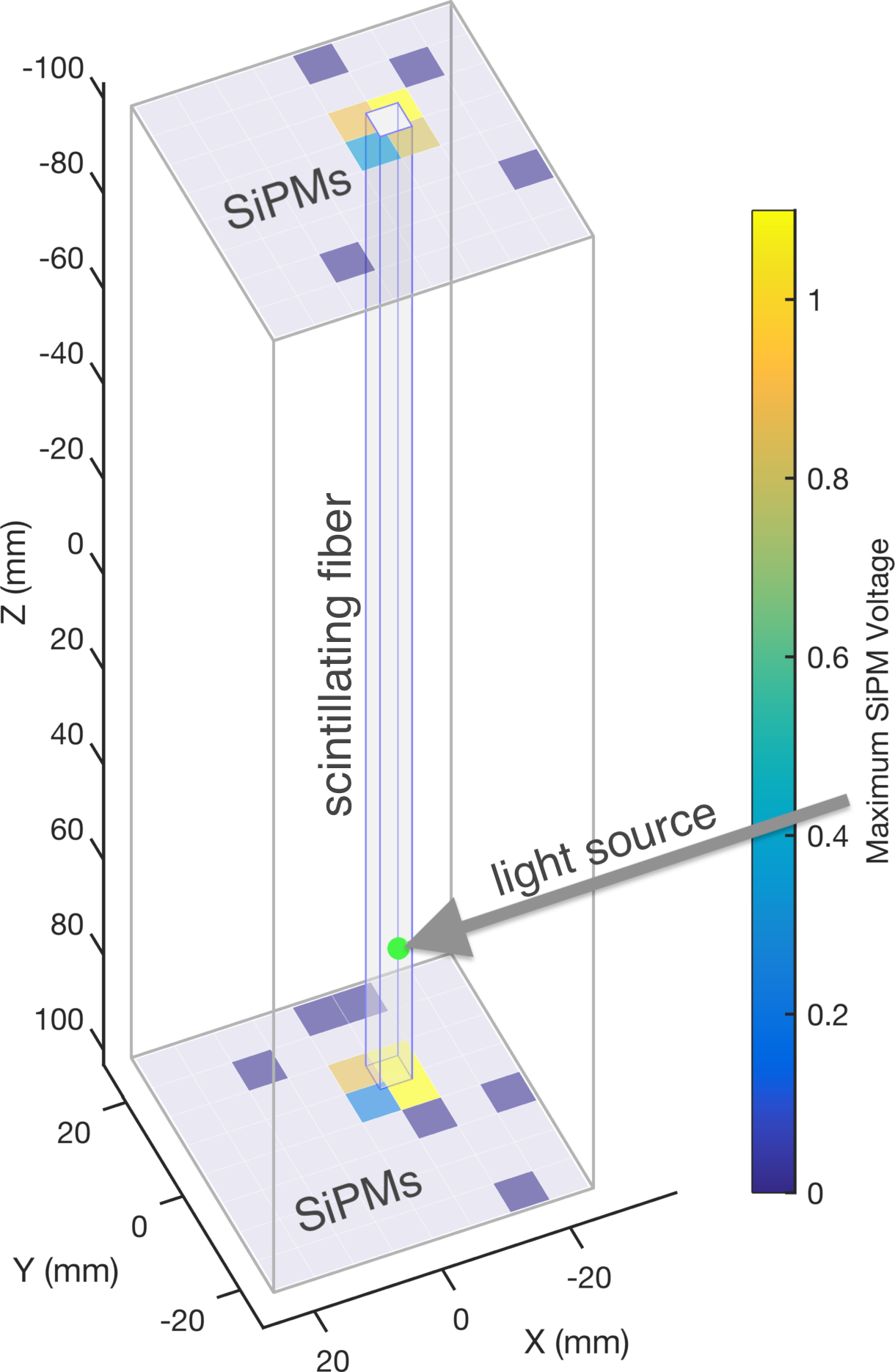}
\caption{TOF-based scintillating-fiber scatter camera. Two SiPM arrays (SensL ArrayJ-60035-64P\cite{SENSLJARRAY}) sandwich a vertical bundle of 5x5 mm square, 20 cm long plastic fibers (one fiber shown). Matching upper and lower SiPM waveform pairs (shown in Figure \ref{f2}) are input to WAVE for scatter reconstruction.}
\label{f0}
\end{figure}

This paper introduces a Machine Learning (ML) Time-Of-Flight (TOF) solution called \textbf{WA}veform \textbf{V}ector \textbf{E}xploitation (\textbf{WAVE}), and evaluates its performance in a common physics setting against traditional TOF techniques. TOF techniques are widely used in the physics community for particle tracking and reconstruction (e.g., \cite{LAPPDtiming,BelleTOF,BESIIITOF}), in stripline anode electronics readouts\cite{anodepaper}, and in the medical industry in Positron Emission Tomography (PET) scanners. Comparisons of different timing techniques relevant to TOF have been performed in the past, in particular by Genat et al.\cite{Genat}, though for only single waveforms, and ML solutions were not included.

Traditional TOF relies on a two step process: first timing waveform arrivals at separate locations, and subsequently reconstructing the source of the waveforms given the measured times. This technique is heavily reliant on accurate timing of the waveform arrivals. In contrast, WAVE directly solves for the waveform's source in a single pass, without the need for any explicitly calculated features. This means that WAVE {\it does not need to solve separately for the arrival time of either waveform}, a novel concept which appears to provide performance benefits in our study.

\section{Time-Of-Flight Methods}

Assuming known propagation speeds $v$ and similar waveform shapes, one may solve for the waveform origin time $t$ and position $x$ via Equations \ref{eq1} and \ref{eq2}, respectively.
\begin{eqnarray}
\label{eq1}
t &=& \frac{t_1+t_2 - L/v}{2}\\
\label{eq2}
x &=& v\left(\frac{t_1-t_2}{2}\right)
\end{eqnarray}

\vspace{2mm}
\noindent where $t_1$ and $t_2$ are the measured waveform arrival times at sides 1 and 2, and $L$ is the distance between measurement locations. $x_0$ is at the center of the fiber, and $-x$ trend towards side 1.

In our example application, Equations \ref{eq1} and \ref{eq2} are applied to the particle scatter camera shown Figure \ref{f0} to reconstruct the scatters within the detector volume. The detector consists of one hundred $5 \times 5 \times 200$ mm$^3$ plastic scintillating fibers arranged in a $10 \times 10$ grid. Gamma or neutron scatters in these fibers create isotropic scintillation light. Due to a high index of refraction mismatch between the fiber cores (n = 1.59) and the small air gaps between the fibers (n = 1.00), much of this light is captured in the fiber cores and reflected to photodetectors at each end, modeled as Silicon PhotoMultipliers (SiPMs)\cite{SENSLJARRAY} for this study. About 25\% of the light produced in the fiber is retained in this manner. The other 75\% attenuates or leaves the detector.

When a photon arrives at one of the SiPM microcells it is converted into an observable current with about 50\% Photon Detection Efficiency (PDE)\cite{SENSLJ}, resulting in a voltage response that is converted to a digital waveform. For our proof of concept simulations, we assume a response consistent with the `fast`' output of commercially available SiPMs, and assume that these voltage waveforms are digitized by DRS4 integrated circuits\cite{DRS4}, which are differentially-driven switched capacitor array ASICs developed by the Paul Scherrer Institut (PSI) in Switzerland. The DRS4 can sample a waveform at up to 5 GigaSamples Per Second (GSPS) with a buffer depth of 1024 samples. Similar switched capacitor array chips include LABRADOR\cite{LABRADOR} and PSEC4\cite{PSEC4}.

\begin{figure}[!htbp]\centering
\includegraphics[width=1\linewidth]{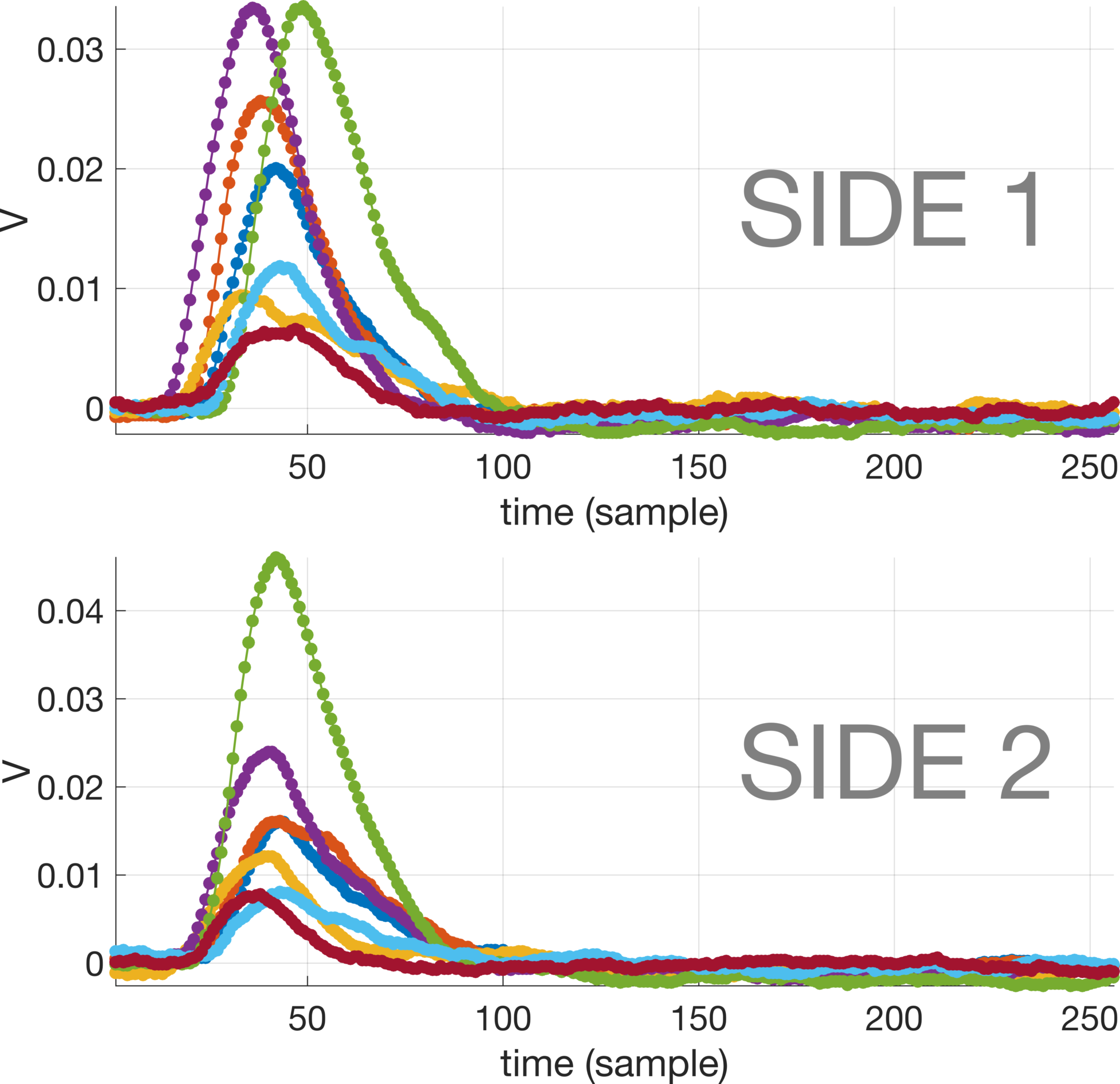}
\caption{Simulated waveform pairs produced by the detector in Figure \ref{f0}. The arrival time difference between the side 1 and side 2 waveforms is used in Equations \ref{eq1} and \ref{eq2} to locate the source time and position along the fiber. WAVE accepts these joined waveforms as inputs of length 256 + 256 = 512.}
\label{f2}
\end{figure}

We use a validated GEANT\cite{GEANT} + MATLAB\cite{MATLAB} MC engine to model our waveforms. This MC package was created by Ultralytics for the mini-TimeCube\cite{mtc} and Large Area Picosecond Photodetector (LAPPD)\cite{LAPPD} collaborations, and allows quick prototyping of new detector designs. Waveforms modeled by this package, along with known truth information are used to analyze the TOF performance of 3 methods:
\begin{enumerate}
    \item Constant Amplitude (CA) TOF
    \item Constant Fraction (CF) TOF
    \item Machine Learning (WAVE)
\end{enumerate}

The MC engine places `perfect' point sources randomly within the scintillating fibers, randomized in time and energy, to create a dataset of 100,000 events, which form the basis for all the results published in this paper. The timing of these point sources was randomized to start anywhere within a 3 ns window, which is a similar trigger jitter to what the DRS4 evaluation board experiences on internally triggered events. The energy of the points was randomized per an exponential distribution with a 250 keVee (keV electron equivalent) exponential constant, to mimic the low energy events a plastic $\sim$MeV neutron scatter camera would likely target.

Digital readout noise was modeled as white and Gaussian with a standard deviation of 0.5 mV. Electronics saturation was modeled at $\pm$ 1.1 V. Single Photo-Electron (PE) waveform amplitudes in our model averaged 0.4 mV $\pm$ 10\% at +5 V SiPM overvoltage. All SiPM characteristics including Photo-Detection Efficiency (PDE), spectral response, dark count rates (80 kHz/mm$^2$ at 20$^\circ$ C) and microcell crosstalk (though not afterpulsing) were modeled per the SensL J-60035 datasheet\cite{SENSLJ}. A Gaussian Transit Time Spread (TTS) of 120 ps $1\sigma$ was modeled for SensL J-60035 PEs as well. 

Optical properties of the scintillating fibers, including scintillation spectrum, index of refraction and yield of 8000 photons/MeV were modeled using a Saint Gobain BCF-10 datasheet\cite{SAINTGOBAIN}. Optical surfaces are assumed to be perfectly flat.

Although we have modeled the full detector, the relevant quantities of comparison for this study are the capability to localize the interaction position along the longitudinal axis of the fiber, as well as the absolute time of the interaction. To that end, the comparisons between the following techniques are done on their performance for a single bar at a time, not including effects of optical crosstalk that might be present in the full detector.

\subsection{Constant Amplitude (CA) thresholding}

\begin{figure}[!htbp]\centering
\includegraphics[width=1\linewidth]{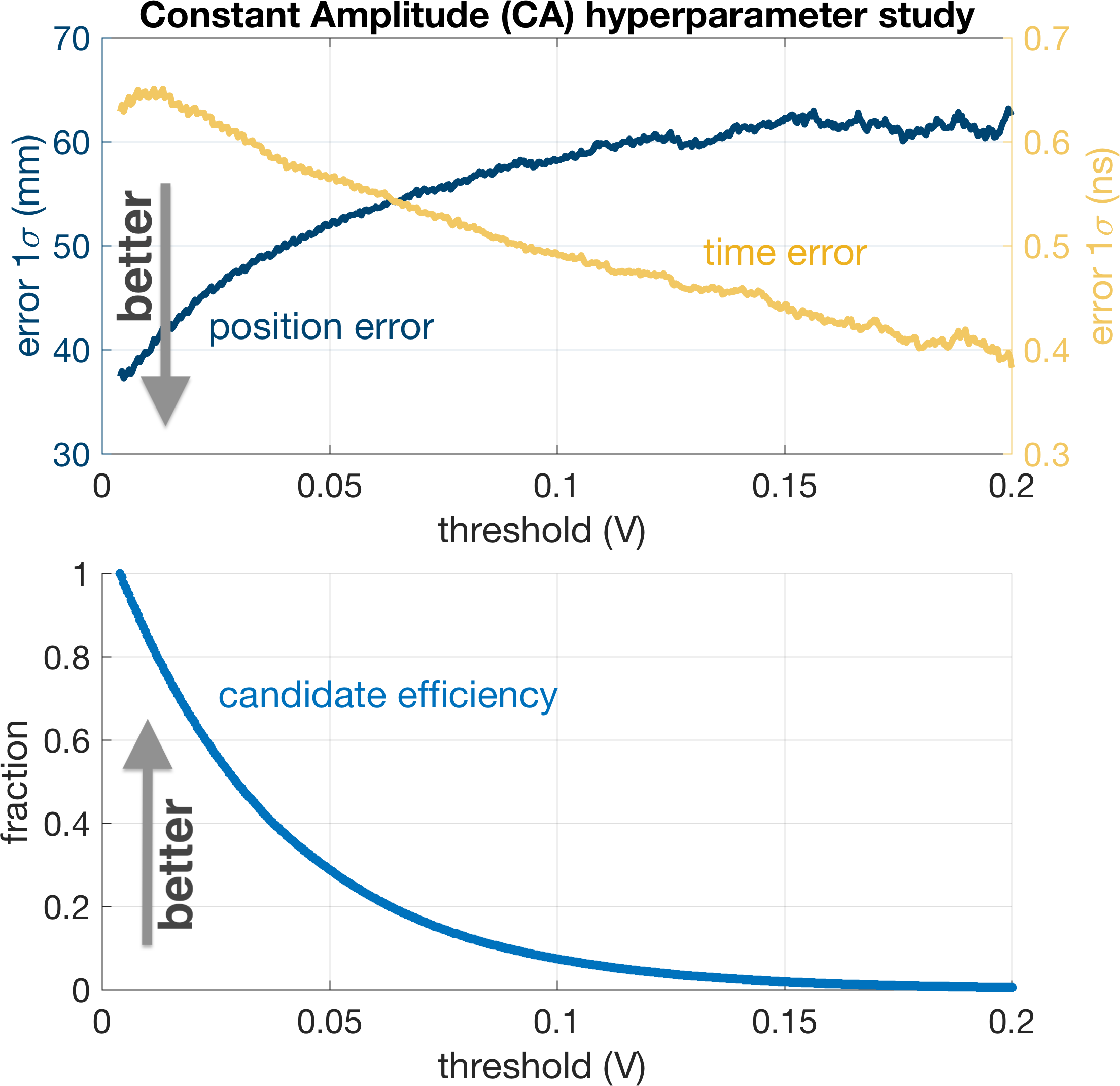}
\caption{CA TOF performance vs amplitude threshold. Amplitudes from 6 - 200 mV are analyzed. Typical waveforms amplitudes are 10-20 mV, as seen in Figure \ref{f2}, but can exceed 500 mV for bright events ($\sim$1 MeVee). The best performance was observed at the lowest threshold (while staying above the noise). Candidate efficiency trends downwards at higher thresholds.}
\label{f5}
\end{figure}

CA thresholding supplies Equation \ref{eq1} with crossing times $t_1$ and $t_2$ extracted from the side 1 and 2 waveforms at the same amplitude. This amplitude is typically set low, in order to extract the very beginning of a waveform risetime, yet high enough to avoid false positives from the noise (which in our model is 0.5 mV 1$\sigma$). Figure \ref{f5} shows a study we performed on this threshold amplitude to determine the optimum value in our application. This search yielded a best amplitude of 6 mV, which we use in reporting our CA results.

\subsection{Constant Fraction (CF) thresholding}

\begin{figure}[!htbp]\centering
\includegraphics[width=1\linewidth]{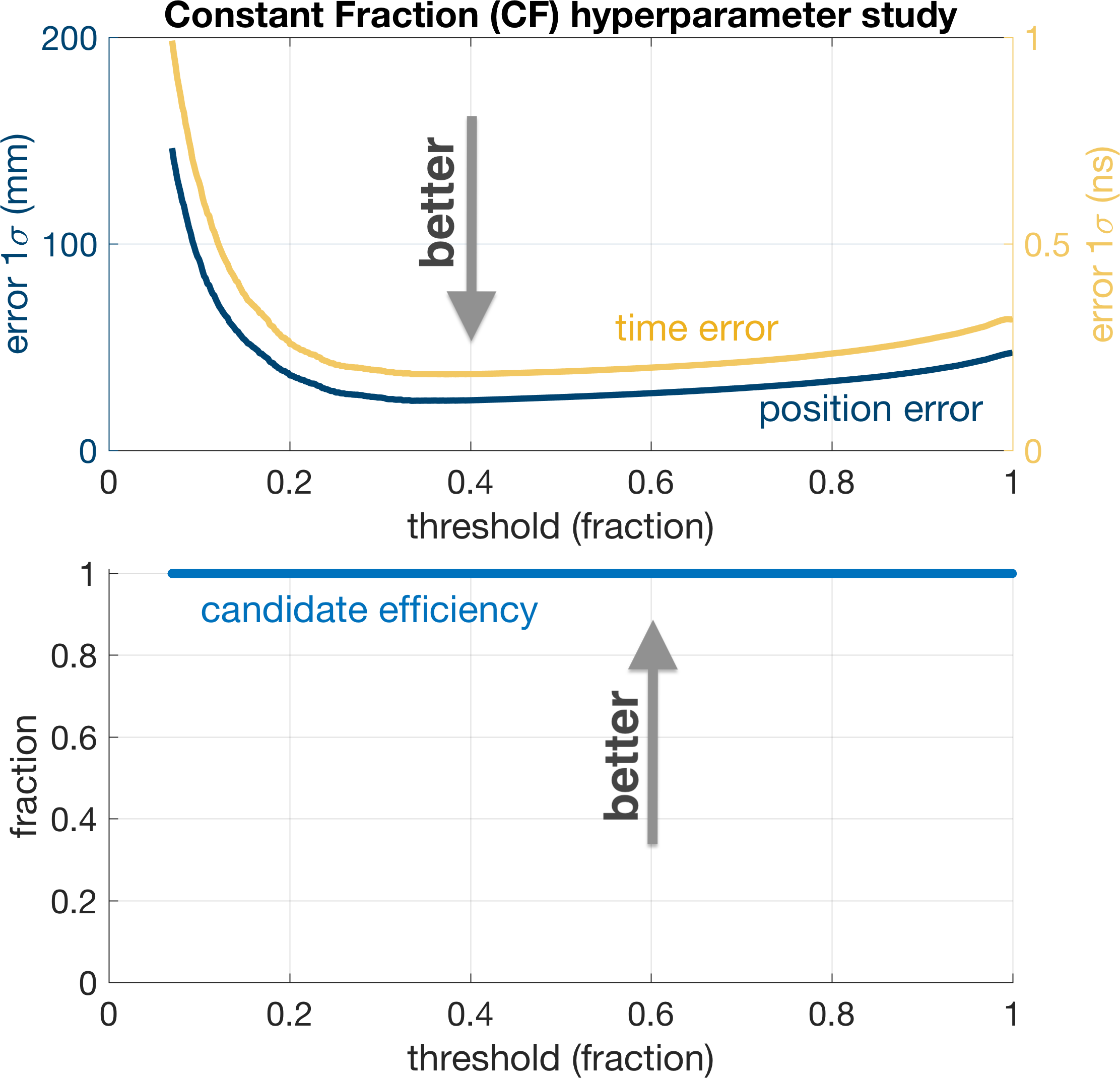}
\caption{CF TOF performance vs fraction threshold. Fractions from 0.0 - 0.99 were analyzed. The best performance was observed around $\sim$0.4.}
\label{f6}
\end{figure}

CF thresholding supplies Equation \ref{eq1} with crossing times $t_1$ and $t_2$ extracted from the waveforms at the same {\it fraction} of max amplitude. This fraction is typically set at 1/2. Figure \ref{f6} shows a study we performed to determine an optimum value for this parameter. The best results are observed around 0.3 - 0.5 fraction. 0.5 is a commonly used, so we will also use 0.5 for reporting CF results.

\subsection{Machine Learning}

\begin{figure}[!htbp]\centering
\includegraphics[width=1\linewidth]{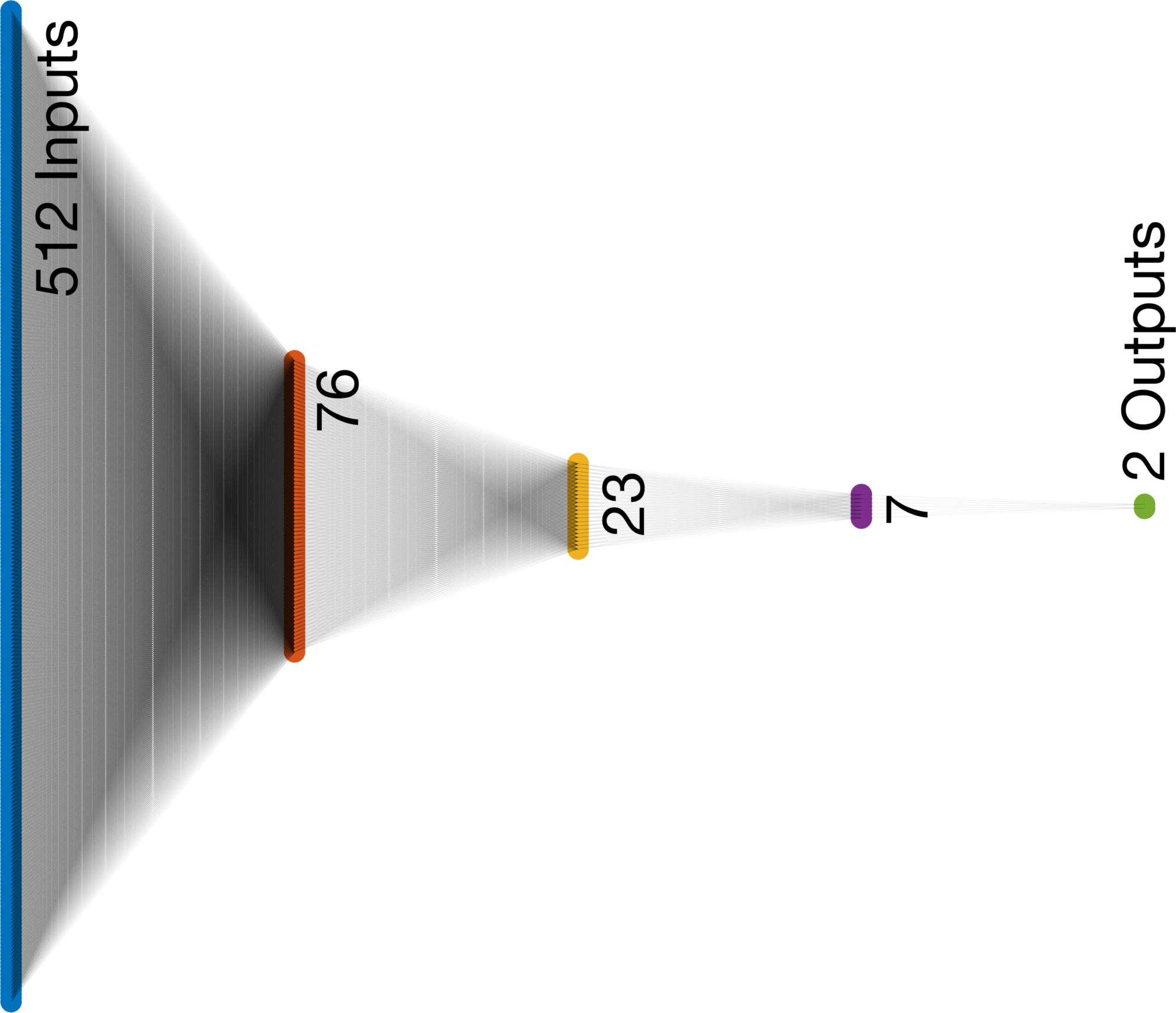}
\caption{WAVE neural network: a 5 layer function-fitting feedforward deep neural network used to reconstruct scatters in the Figure \ref{f0} detector. 512 inputs (the joined waveforms shown in Figure \ref{f2}) pass through 3 fully-connected hidden layers (weights + bias + $tanh()$ activation). The 2 outputs are scatter position $x$ (mm) and time $t$ (ps) in the detector.}
\label{f3}
\end{figure}

WAVE is a 5-layer function-fitting feedforward network, shown in Figure \ref{f3}. We constructed WAVE in Python using the PyTorch 0.4.0\cite{pytorch} package, and made it freely available under the open source GPL-3.0 license at \url{https://github.com/ultralytics/wave}. A 512-length input layer, composed of two joined waveforms, connects to 3 hidden layers of size 76, 23 and 7, with an output layer of size 2: 1 position $x$ and 1 time $t$. Side 1 and Side 2 waveforms shown in Figure \ref{f2} of length 256 each are joined sequentially to create each 512-length input vector. 

All hidden layer neurons are linear (weights + bias), with $tanh()$ activations, except layer 4, which lacks activation. This structure was settled on using trial and error, though the exact geometry is not critical, as we observed that our Section 3 results were robust to moderate changes in network geometry. Our 100,000 event dataset was split into 3 groups:

\begin{enumerate}
\item {\bf Train (70\%).} Used during WAVE training to update network parameters (weights and biases).
\item {\bf Validate (15\%).} Used to stop WAVE training to prevent over-training. Not used to update network parameters.
\item {\bf Test (15\%).} Used to evaluate performance of CA, CF and WAVE. Not used during WAVE training.
\end{enumerate}

We used an Adam optimizer with near-default hyperparameters ($lr=0.001$, $\epsilon=1\times 10^{-8}$, $\beta_1=0.9$, $\beta_2=0.999$) and a Mean Squared Error (MSE) loss criterion. Supervised learning (on the Train and Validate datasets) was performed on a Google Cloud Platform (GCP) Compute Engine Virtual Machine (VM) instance with one Nvidia Tesla P100 GPU, taking $\sim$ 5 minutes of wall-clock time for 50,000 epochs, and yielding a minimum validation loss of 0.023 after 14,278 epochs. The epoch 14,278 model was frozen for inference on the Test dataset to produce the Section 3 results.

\section{Results}

\begin{figure}[!htbp]\centering
\includegraphics[width=1\linewidth]{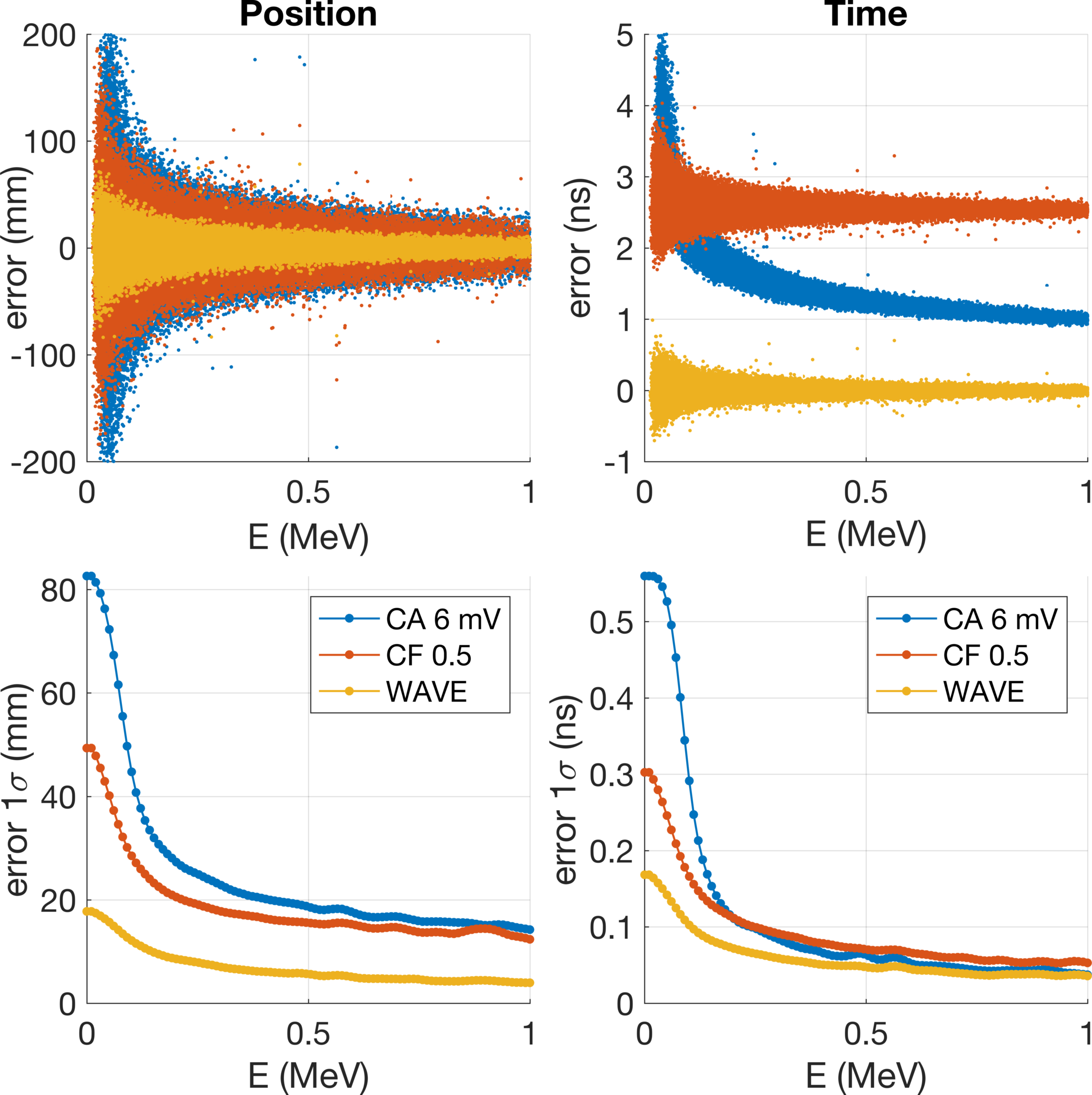}
\caption{Results for CA, CF, and WAVE vs scatter energy E. {\bf TOP:} Reconstruction errors. {\bf BOTTOM:} Reconstruction error 1$\sigma$, averaged over energy in Table \ref{table:1} (biases are ignored in our results, though we note WAVE automatically learns and eliminates these). WAVE performance exceeds CA and CF for both metrics across all energies in our dataset.}
\label{f4}
\end{figure}

Results are summarized in Table \ref{table:1} and detailed in Figure \ref{f4} as a function of scatter energy. WAVE performance exceeds CA and CF in our study by a wide margin in both position and time reconstruction. {\bf The largest performance increases are observed at the lowest energy events}, the most challenging region of the parameter space to reconstruct. 

\begin{table} [!htbp]
\fontsize{10}{10} \selectfont \centering
\begin{tabular}{l|lllll}
method		& candidates 	& position $x$		& time $t$			\\
        	& \bf fraction	& \bf mm 1$\sigma$ 	& \bf ps 1$\sigma$ 	\\
\hline 	
CA - 6 mV	& 0.95 			& 38     			& 627 				\\
CF - 0.5	& 1.00 			& 26       			& 150				\\
\bf WAVE	& \bf 1.00		& \bf 11		    & \bf 90			\\
\end{tabular}
\caption{MC study results (lower 1$\sigma$ is better), showing WAVE exceeding CA and CF performance in both spatial and temporal reconstruction. This table summarizes Figure \ref{f4}, which plots the above metrics vs scatter energy.}
\label{table:1}
\end{table}

\section{Conclusion}

Table \ref{table:1} and Figure \ref{f4} show clear performance increases for WAVE compared to traditional techniques across all regions of the parameter space evaluated in our study.

The next steps are the application of this technique to real-world data. Challenges to real-world implementation include the need to accumulate training data of sufficient {\it quantity} and {\it variation} to generalize well in real-world scenarios. For applications where such training data it attainable, it appears WAVE may offer significant performance benefits over traditional solutions.

\section{Acknowledgements}
This effort was initiated and funded by Ultralytics LLC (\url{https://www.ultralytics.com}) in collaboration with the University of Hawaii and Lawrence Livermore National Laboratory. Lawrence Livermore National Laboratory is operated by Lawrence Livermore National Security, LLC, for the U.S. Department of Energy, National Nuclear Security Administration under Contract DE-AC52-07NA27344. Release LLNL-JRNL-761634.

\appendix
\section{}

\begin{figure}[!htbp]\centering
\includegraphics[width=1\linewidth]{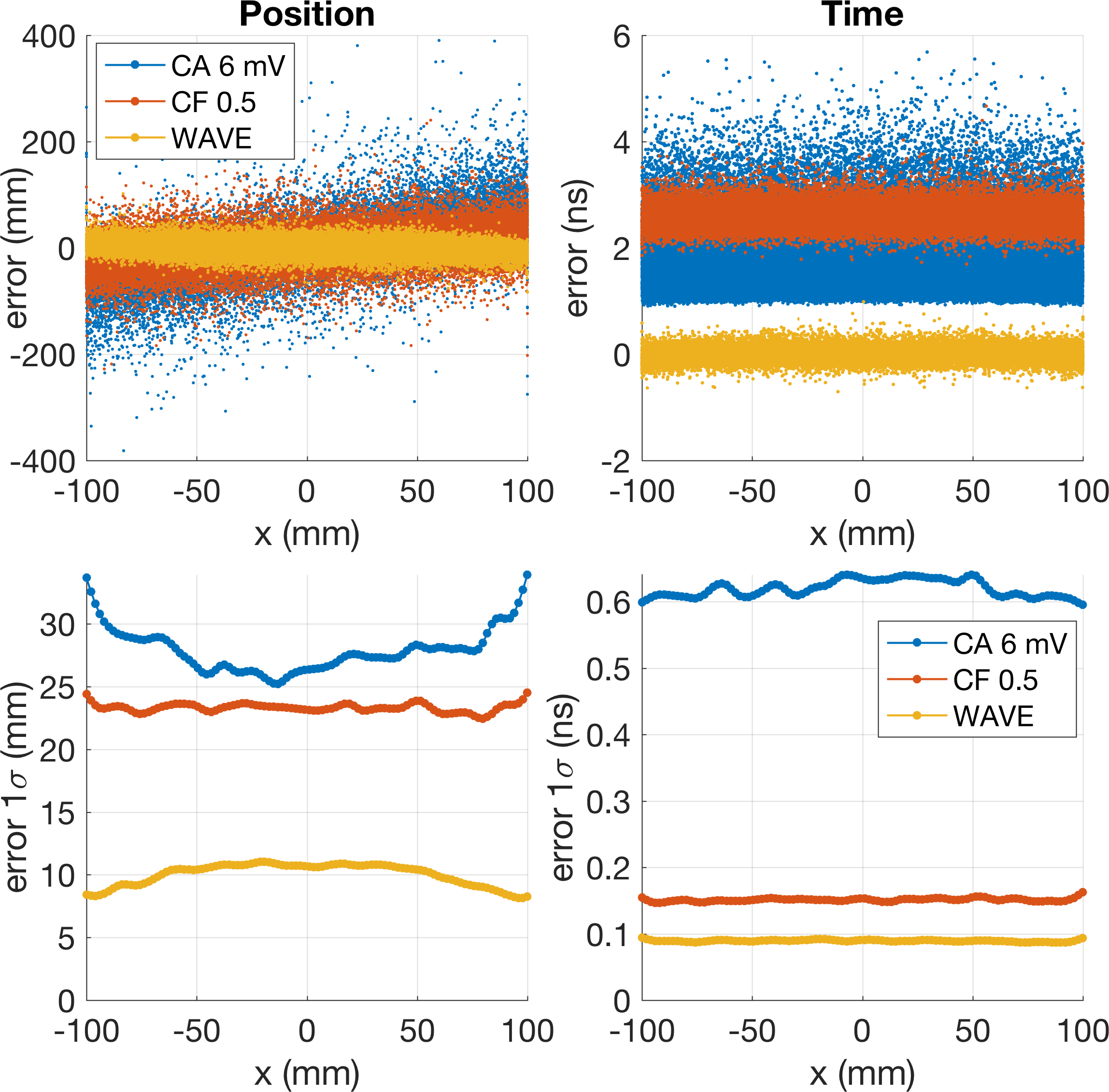}
\caption{Results vs scatter {\bf position} (in fiber). Edge affects appear in all 3 fitting techniques, though interestingly WAVE is the only technique to show improved performance near the fiber edges. WAVE outperform CA and CF across all regions of this parameter space (scatter position space), as it does in Figures \ref{f4} and \ref{fa1}.}
\label{fa0}
\end{figure}

\begin{figure}[!htbp]\centering
\includegraphics[width=1\linewidth]{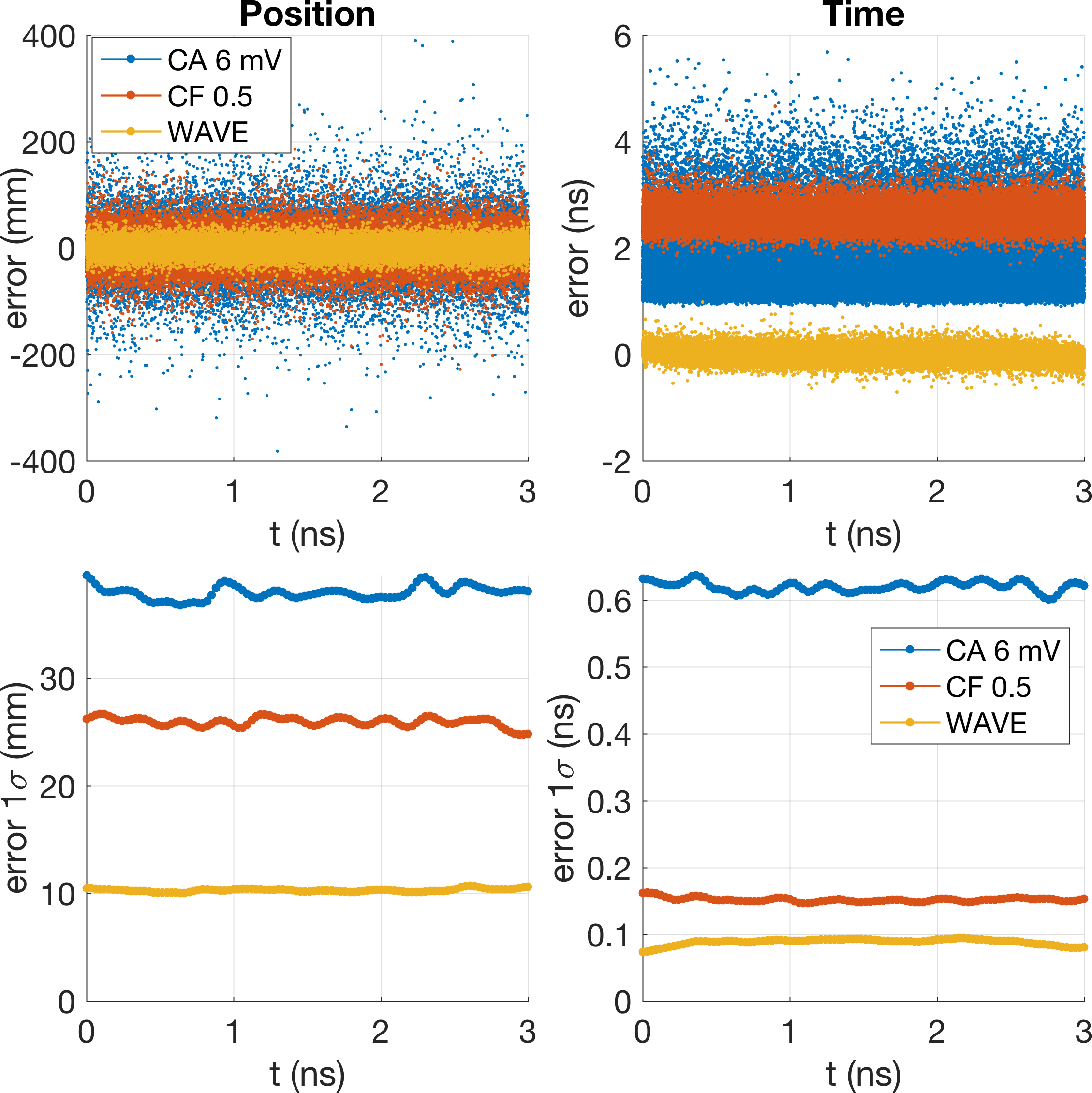}
\caption{Results vs scatter {\bf time}. No significant edge effects are observed. WAVE outperform CA and CF across all regions of this parameter space (scatter time space), as it does in Figures \ref{f4} and \ref{fa0}.}
\label{fa1}
\end{figure}

\begin{figure}[!htbp]\centering
\includegraphics[width=1\linewidth]{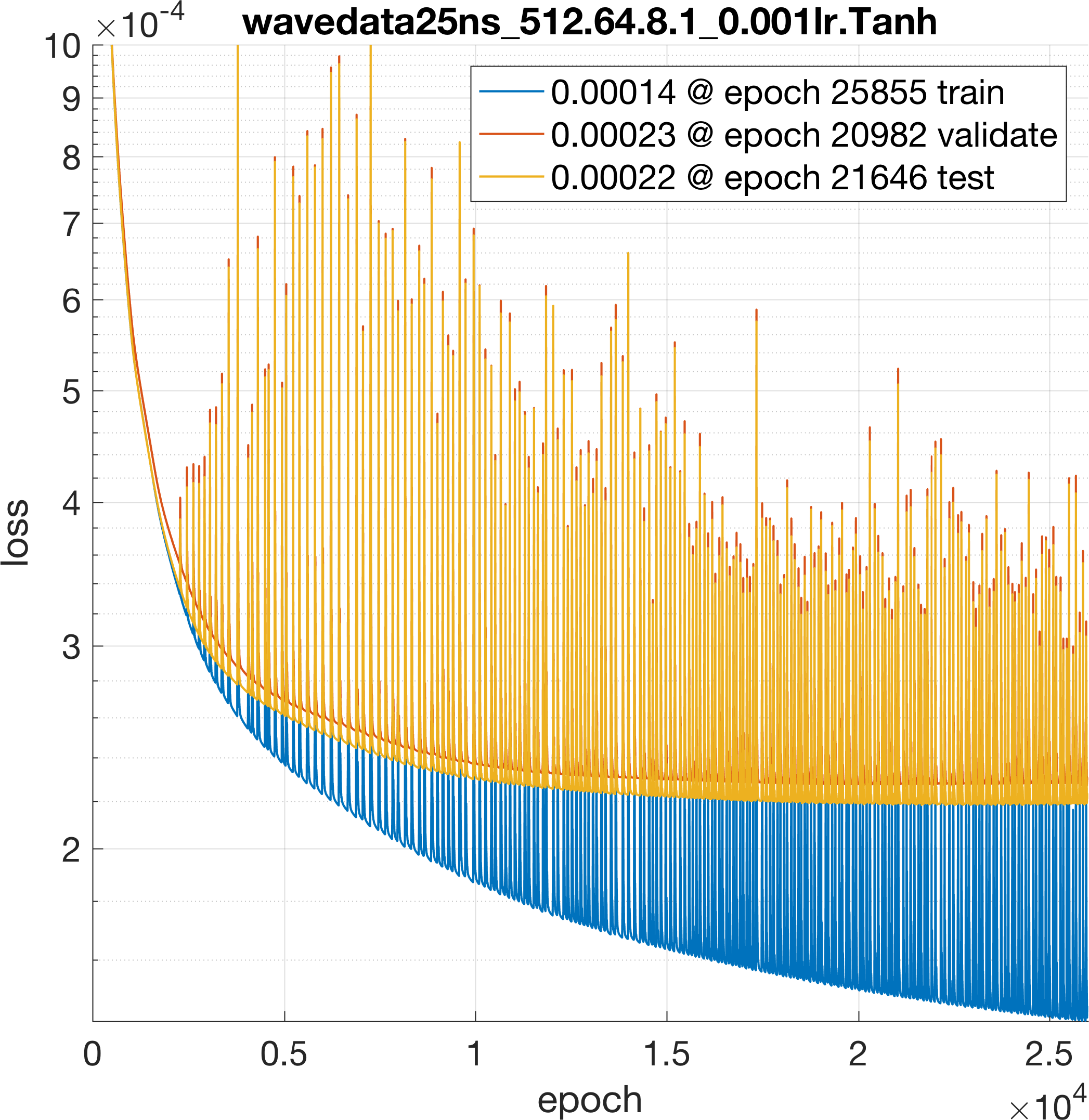}
\caption{WAVE training loss vs number of epochs trained. Validation and Test losses reach a minimum around 20,000 epochs, after which they begin to increase. Test losses continue to decrease. Training is terminated after 5,000 validation epochs fail to register a new absolute minimum.}
\label{fa2}
\end{figure}

\begin{figure}[!htbp]\centering
\includegraphics[width=1\linewidth]{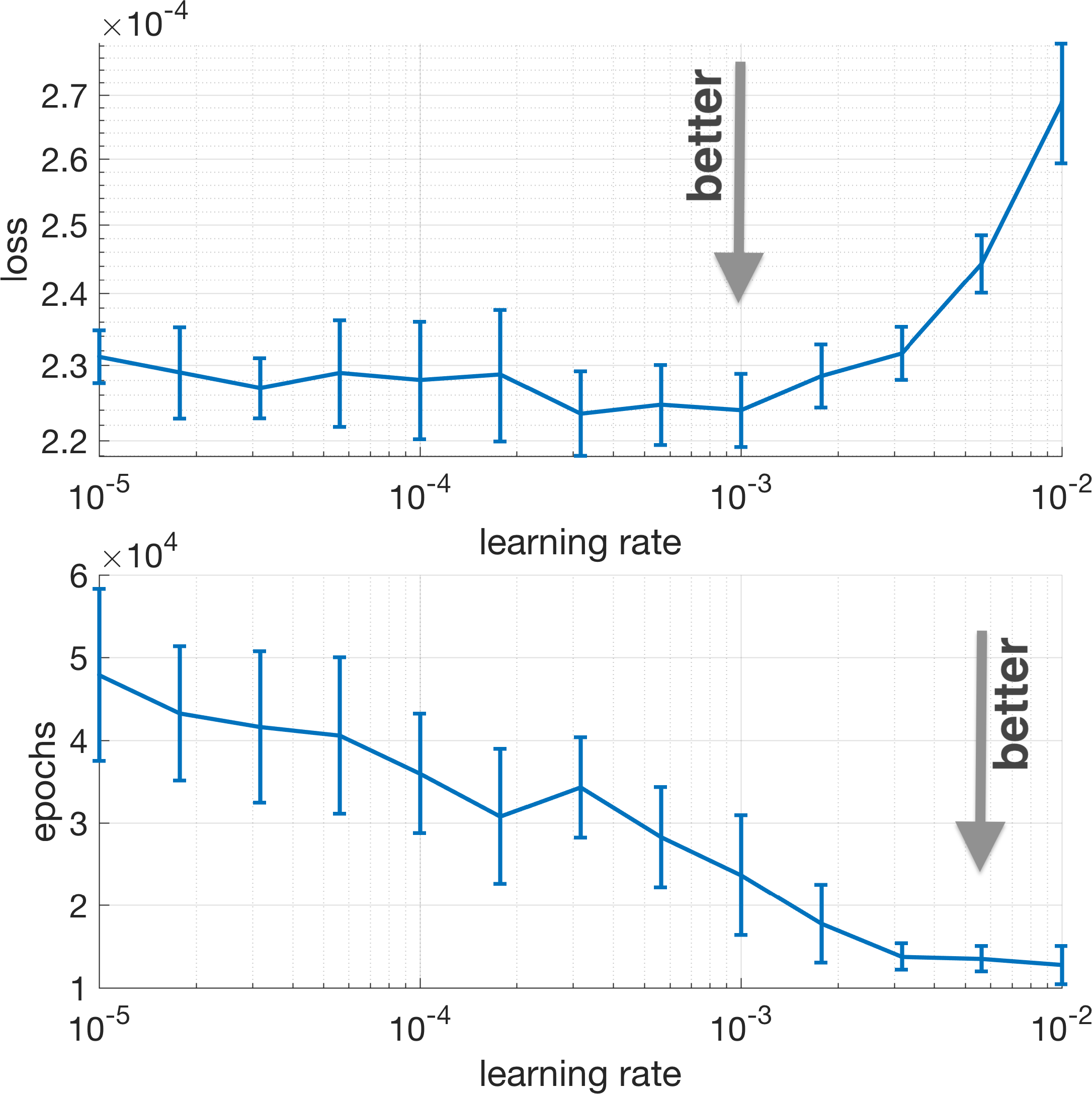}
\caption{Hyperparameter study: effect of learning rate on best validation loss and epochs required to reach best validation loss. The default learning rate of 0.001 appears to be well placed, as it allows for a near optimal final loss in a relatively short training time.}
\label{fa4}
\end{figure}

\begin{figure}[!htbp]\centering
\includegraphics[width=1\linewidth]{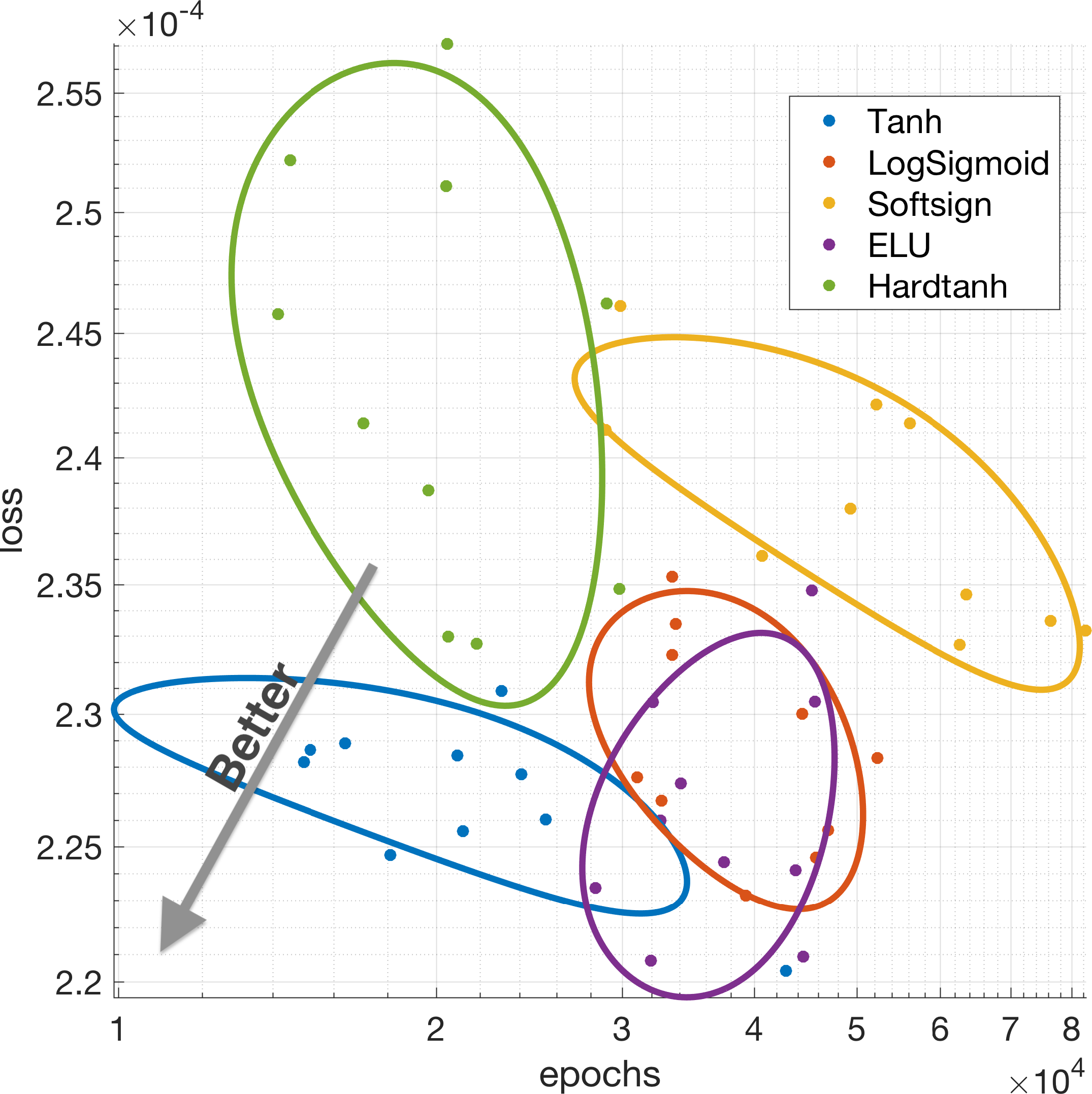}
\caption{Activation function study. 5 activation function candidates are evaluated for use in WAVE. ELU and Tanh both appear well suited to the problem. Softsign produces the worst results.}
\label{fa5}
\end{figure}

\begin{figure}[!htbp]\centering
\includegraphics[width=1\linewidth]{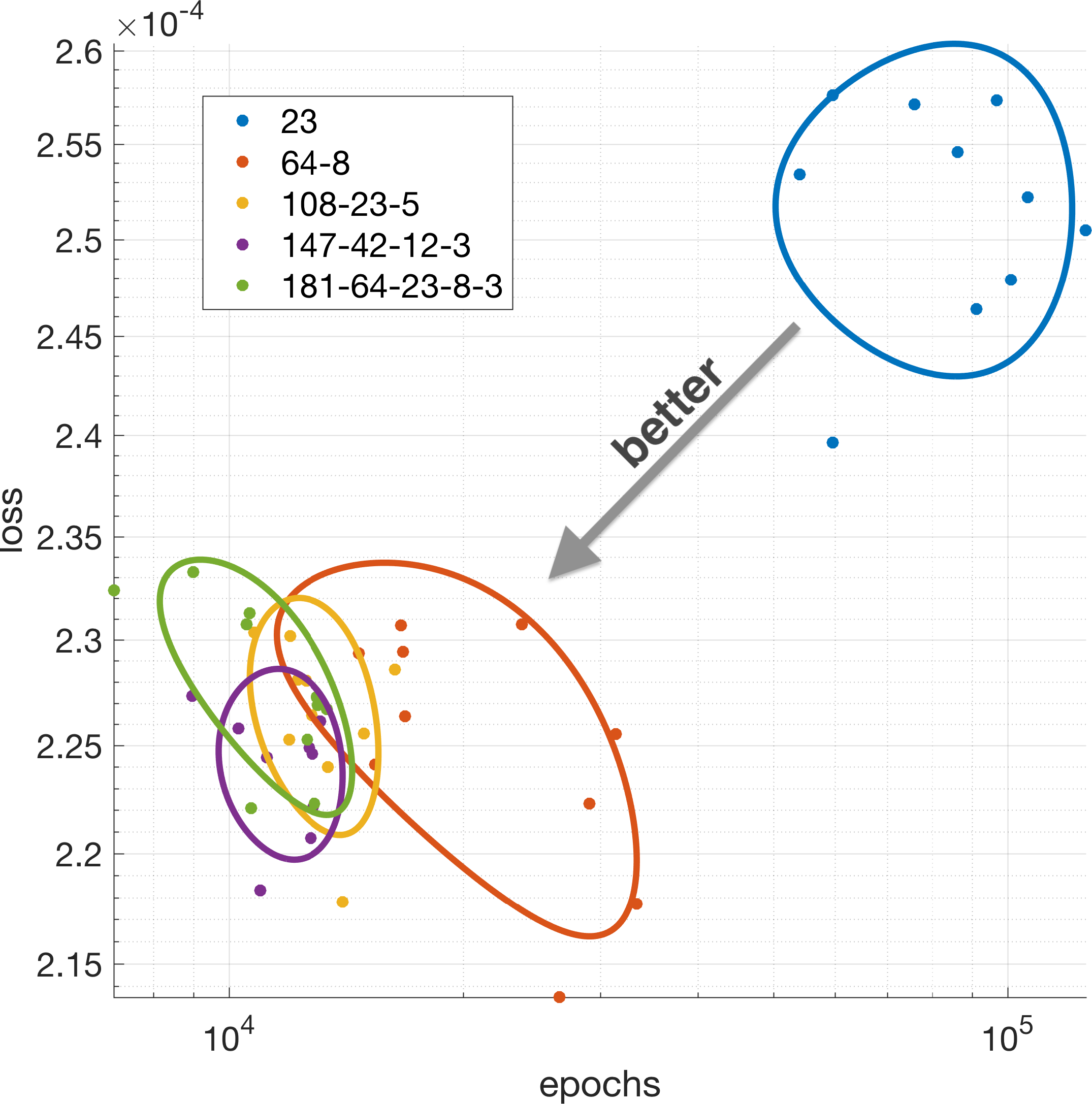}
\caption{Hidden layer structure study. Different hidden layer structures were analyzed, starting at 1 hidden layer with 23 neurons and moving to 5 hidden layers with 181 neurons in the first layer, 64 in the second layer, etc. The largest performance increase was observed when moving from 1 to 2 layers, with corresponding training speed improvements as well, yet with diminishing returns beyond 3-4 layers.}
\label{fa6}
\end{figure}

\begin{figure}[!htbp]\centering
\includegraphics[width=1\linewidth]{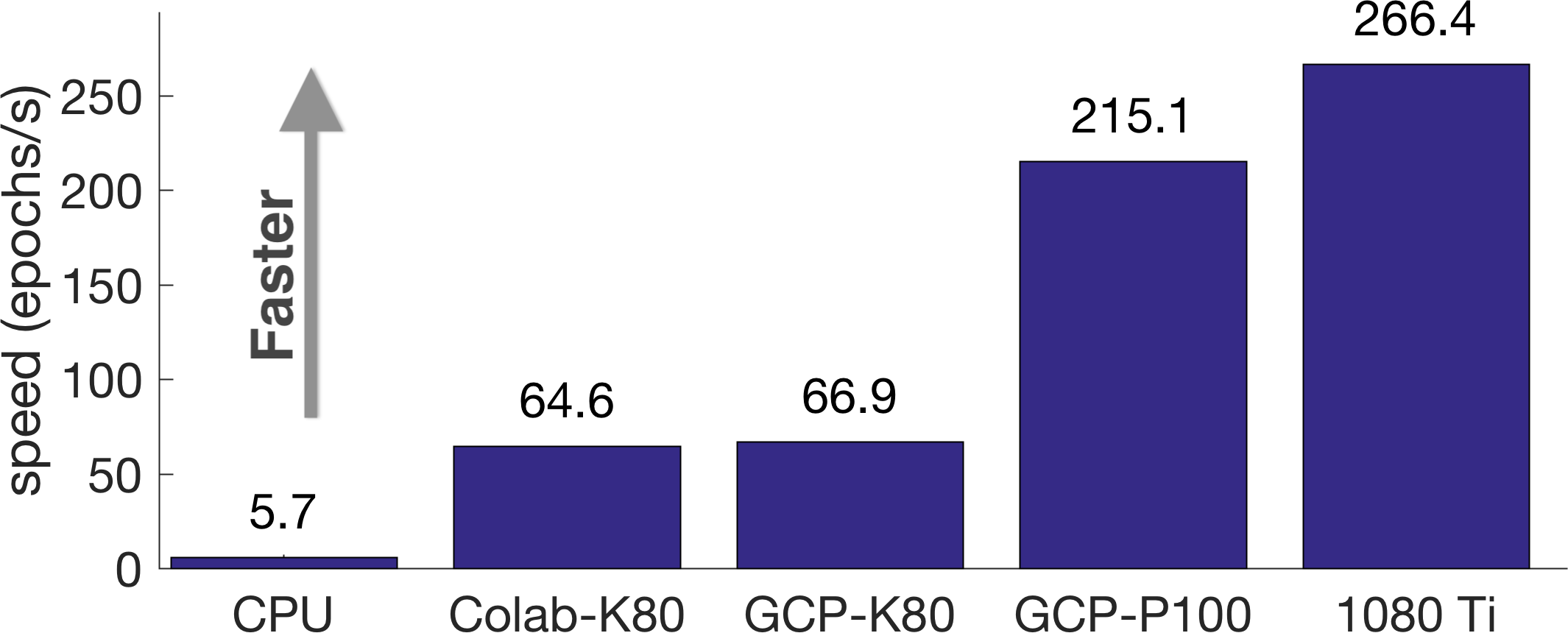}
\caption{WAVE pytorch training speed on various hardware, measured in epochs per second. About 20,000 epochs are required for training, corresponding to $\sim$ 100 seconds on a Nvidia Tesla P100 GPU.}
\label{fa3}
\end{figure}

\end{document}